# Dimensional cross-over of the charge density wave order parameter in thin exfoliated 1T-VSe$_2$

Árpád Pásztor, Alessandro Scarfato, Céline Barreteau, Enrico Giannini, and Christoph Renner*

*DQMP,Université de Genève, 24 quai Ernest Ansermet, CH-1211 Geneva, Switzerland.*

**The capability to isolate one to few unit-cell thin layers from the bulk matrix of layered compounds[1] opens fascinating prospects to engineer novel electronic phases. However, a comprehensive study of the thickness dependence and of potential extrinsic effects are paramount to harness the electronic properties of such atomic foils. One striking example is the charge density wave (CDW) transition temperature in layered dichalcogenides whose thickness dependence remains unclear in the ultrathin limit[2-5]. Here we present a detailed study of the thickness and temperature dependences of the CDW in VSe$_2$ by scanning tunnelling microscopy (STM). We show that mapping the real-space CDW periodicity over a broad thickness range unique to STM provides essential insight[6]. We introduce a robust derivation of the local order parameter and transition temperature based on the real space charge modulation amplitude. Both quantities exhibit a striking non-monotonic thickness dependence that we explain in terms of a 3D to 2D dimensional crossover in the FS topology. This finding highlights thickness as a true tuning parameter of the electronic ground state and reconciles seemingly contradicting thickness dependencies determined in independent transport studies.**

Following the ground-breaking exfoliation of graphite into one atom thin carbon sheets, an increasing number of layered compounds can now be isolated from their bulk matrix in the form of one to few unit-cell thin layers. These sheets often feature unique[7-9] or enhanced[2,10] properties in comparison to their parent bulk compounds. They depend on material thickness and can be further tuned through doping, electrostatic gating and assembly of distinct layers into complex heterostructures. Transition metal dichalcogenides (TMDs) are of particular interest in this context. They can be readily exfoliated into thin flakes down to the single unit-cell limit[11] and offer a unique playground for studying the thickness dependence of their electronic properties. For example, in MoS$_2$, photo active transitions become available in the single layer limit due to the appearance of a direct gap in the band structure. Metallic TMDs host a variety of electronic phases like superconductivity and charge density waves (CDW)[12-17] whose transition temperatures can be modified by reducing the thickness of the host crystal[2,4,5,10]. Exfoliation offers a new degree of freedom to engineer these electronic ground states. Unfortunately, thin flakes of metallic TMDs often degrade in air[18] and have thus been much less studied than their semiconducting and insulating counterparts. To overcome this limitation, we developed a mechanism enabling in-situ exfoliation.

This study is focused on the thickness dependence of the CDW phase in VSe$_2$, a metallic TMD that grows in the $1T$ polymorph. It consists of van der Waals bonded slabs of triangular vanadium layers sandwiched between two triangular selenium layers (Fig. 1a, left inset). Each vanadium is surrounded by six selenium atoms in an octahedral configuration with in-plane and out of plane lattice constants $a = b = 3.36$ Å and $c = 6.104$ Å, respectively[19]. Bulk $1T$-VSe$_2$ undergoes a CDW phase transition at $T_c^{bulk} \simeq 105$ K into a commensurate $4a \times 4a$ superlattice within the layers (ab-plane) and an incommensurate $\sim 3.1c$ modulation along the c-axis[19-21]. The CDW transition temperature in thin flakes ($T_c$) has been found to deviate up to 30% from these bulk values, with contradicting findings where $T_c$ is either increased[4] or reduced[5], seemingly dependent on sample preparation.

Before investigating the thickness dependence of the CDW, we have characterized it in bulk single crystals by transport measurements and STM. Resistivity as a function of temperature (Fig. 1a) shows a characteristic kink at the CDW phase transition $T_c^{bulk} \simeq 105$ K, in agreement with previous studies[22]. Constant current STM images and corresponding Fourier transforms at 77.6 K (Figs. 1c,d) clearly reveal a triangular atomic lattice ($a = 3.36$ Å) and the in plane $4a \times 4a$ commensurate CDW modulation[23]. Tunneling spectroscopy (Fig. 1b) is consistent with data reported elsewhere[24], including a characteristic conductance peak associated with the vanadium derived $d$-band below the Fermi level ($E_F$) and an asymmetric $U$-shaped background centered on $E_F$.

To gain insight into the thickness dependence of the CDW, we take advantage of steps and terraces naturally present on the exfoliated flakes. The local thickness is directly quantified from the STM topographic traces as the height of the terrace above the reconstructed Au(111) substrate. Here, we concentrate on STM micrographs measured above 77 K, near $T_c^{bulk}$ where thickness dependent CDW features are most prominent. Topographic and CDW features imaged by STM on different exfoliated thin flakes and terraces with distinct thicknesses (Fig. 2) are very similar to those in bulk crystals. Remarkably, we observe the same $4a \times 4a$ charge order down to the thinnest sample studied (2.2 nm). However, a closer inspection of the 77.6 K micrographs reveals a noticeably weaker CDW amplitude in the 20 nm thin region than in all other thicknesses. At 95.0 K, closer to $T_c^{bulk}$, the CDW is almost completely suppressed on the 20 nm and 50 nm thin flakes - similar to what we observe in bulk crystals - while it remains surprisingly strong on the thinnest 10 nm flake.

A more quantitative analysis of the CDW contrast revealed by STM is required to go beyond the above approximate visual assessment. The amplitude of the gap near $E_F$ associated with the CDW phase transition would be a natural order parameter. However, tunneling spectroscopy does not show any significant reduction in the LDOS at $E_F$ at the phase transition (Fig. 1b), consistent with the tiny portion of the Fermi surface that is gaped in the CDW phase of $1T$-VSe$_2$ [25]. Moreover, the vanadium derived $d$-band just below $E_F$ obscures the CDW gap.

A convenient alternative measure to describe the phase transition when the quasi-particle gap is not clearly observable is the CDW modulation amplitude[26], which depends linearly on the gap in a mean field description[27]. For a quantitative analysis of the CDW phase transition, we thus introduce an order parameter $\psi$ defined as

$$\psi = \frac{\int_{S_{CDW}} I(k_x,k_y) \mathrm{d}k_x \mathrm{d}k_y}{\int_{S_{lattice}} I(k_x,k_y) \mathrm{d}k_x \mathrm{d}k_y} \qquad (1)$$

where $I(k_x, k_y)$ is the amplitude in Fourier space and $S_{CDW}$ and $S_{lattice}$ are circular shaped integration areas around the CDW and lattice peaks, respectively (Fig. 1d). $S_{CDW}$ and $S_{lattice}$ were chosen such that the k-space area is the same for all examined micrographs ($|S_{CDW}| = |S_{lattice}| = 0.2$ nm$^{-2}$). We normalize to the atomic lattice components to account for possible differences in tunneling conditions, which result in small variations in the appearance of the CDW patterns imaged by STM. Note this variability is the same for bulk crystals and thin exfoliated flakes and does not affect the calculated order parameter $\psi$.

The temperature dependence of $\psi$ near the phase transition of a single crystal (Fig. 3a) can be modeled in a mean field description by a phenomenological BCS gap equation[28]:

$$\psi(T) = A \cdot T_c \cdot \tanh\left(1.74 \cdot \sqrt{\frac{T_c}{T} - 1}\right) \qquad (2)$$

where $A$ is a scaling factor to be determined. The solid line in Fig. 3a is a fit of equation (2) to the experimental data points, where we have set $T_c = T_c^{bulk} = 105$ K. The fit has a single adjustable parameter which we find to be $A = 0.0165 \pm 0.0004$. The consistent picture emerging from this analysis confirms that equation (1) is an adequate measure of the order parameter enabling a quantitative analysis of the CDW phase transition. Examining a large set of STM images from different terraces and flakes at $T = 77.6$ K, we find a non-monotonic thickness dependence of the CDW order parameter (Fig. 3c). $\psi$ is gradually decreasing from its bulk value when reducing the crystal thickness down to 20 nm. When thinning the crystal further, below 20 nm, the thickness dependence is reversed and $\psi$ is increasing to even significantly exceed the bulk value in the thinnest regions measured here (2.2 nm).

The order parameter $\psi$ is not suitable for a direct comparison of our STM data with published transport experiments[4,5]. The latter report $T_c$ as a function of thickness, which is challenging to measure by STM because of thermal drift making it difficult to maintain the tip position over a specific location while changing the temperature above 77 K. As it turns out, we can use equation (2) to calculate the local $T_c$ based on $\psi(T)$ extracted from STM images at a given temperature $T < T_c$. Assuming the scaling factor $A$ is the same for all thicknesses, equation (2) provides a direct correspondence between $T_c$ and the order parameter $\psi(T)$. To verify this assumption, we plot equation (2) for the 10 nm, 20 nm, and 50 nm thin terraces, using the transition temperature $T_c$=122 K, 87 K, and 100 K, respectively, calculated for each thickness from $\psi(T = 77.6\ K)$. The agreement with the experimental data points is excellent as shown in Fig. 3b. This demonstrates the validity of this method, providing an unprecedented ability to determine the local CDW transition temperature solely based on the charge modulation amplitude measured by STM. The expected CDW gap amplitudes within this weak coupling model are in the range of 3.8 meV to 5.7 meV depending on crystal thickness, too small to be properly resolved above 77 K.

The CDW transition temperature calculated for different thicknesses using the above method are plotted in Fig. 4. They are in remarkable quantitative agreement with independent transport studies[4,5]. The most striking findings of our analysis are a non-monotonic thickness dependence of $T_c$ and a significant increase of $T_c$ above the bulk value in the thinnest terraces measured here. The opposite thickness dependencies of $T_c$ we find in very thin compared to thicker terraces lift the contradicting results reported by Xu et al.[4] and Yang et al.[5]. That discrepancy has been blamed on the distinct liquid and mechanical exfoliation techniques used in these studies, when it is in fact the result of investigating different thickness ranges. The STM data presented in Fig. 4 suffers no such ambiguity; they were obtained with the same tip probe on flakes prepared in an identical mechanical exfoliation process.

The two distinct and opposite thickness dependencies of the CDW transition temperature shown in Fig. 4 suggest a crossover from a three-dimensional (3D) to a two-dimensional (2D) regime around 20 nm. $T_c$ has been found to increase with decreasing thickness also for other very thin TMD compounds[2,10]. However, these studies lack important real space information to fully assess the nature of the CDW. STM directly and unambiguously shows no alteration in the CDW symmetry and periodicity with thickness and temperature. The only modification we observe is the charge modulation amplitude associated with the change in $T_c$. We propose that the enhanced $T_c$ in the thinnest samples is a consequence of spatial confinement, in analogy to BCS superconductors governed by a gap equation similar to equation (2). In that case, for a confinement potential $U$ above a critical value, theory predicts the superconducting transition temperature ($T_{sc}$) to increase with decreasing thickness before vanishing to zero in the zero thickness limit[29]. The characteristic thickness $d_{max}$ for which $T_{sc}$ is maximum in this model depends on $U$, on the coupling strength, and on the carrier density. It is important to note that $d_{max}$ does not necessarily correspond to the single layer limit. It can be larger and it is thus necessary to examine a range of thicknesses to draw definite conclusions about the thickness dependence of $T_c$. In particular, considering only bulk and single layer crystals may lead to contradicting conclusions about the effect of dimensional confinement, even in the same material if different preparation and substrates result in different $d_{max}$.

The decreasing $T_c$ with decreasing thickness above 20 nm can be understood considering the Fermi surface (FS) topology of VSe$_2$. It has a significant dispersion of a few eV along $k_z$[21, 25], different from the mostly 2D FS of other layered TMDs. Photoemission[21,30] reveals large parallel FS portions centered at the M(L) points of the Brillouin zone. They offer good in-plane nesting conditions that persist for all $k_z$. This nesting is strongest at a particular $k_z$, resulting in an effective out of plane nesting vector and a 3D CDW[21] in bulk VSe$_2$. Upon thinning the crystal to 56 nm and 20 nm (~93 and ~33 layers), the out-of-plane nesting condition becomes weaker due to the discretization of the FS by the reduced number of available $k_z$ points. This drives the system into a weaker 2D charge order that is further suppressed by enhanced fluctuations expected in 2D.

In summary, we find a striking non-monotonic thickness dependence of the CDW transition temperature in mechanically exfoliated 1$T$-VSe$_2$ from bulk to 2.2 nm thin flakes. On the other hand, the modulation period and alignment with the atomic lattice are entirely independent on thickness. We propose this behavior is a consequence of a 3D to 2D dimensional crossover in the FS topology around 20 nm thickness combined with quantum confinement in thinner flakes. The dimensional crossover weakens the bulk CDW in the thicker flakes and the confinement enhances it in the thinner ones. Unambiguous evidence for this behavior is provided by the local $T_c$ determined from the CDW order parameter measured by STM over an unprecedented broad range of thicknesses in a given experiment. We demonstrate that the charge modulation amplitude provides a suitable measure of the CDW phase transition order parameter. This allows a robust determination of the local critical temperature based solely on STM topographic images of the CDW at a given temperature below $T_c$. Interestingly, the exact same approximate form of the (weak coupling) BCS equation quantitatively describes the CDW order parameter and critical temperature, independent of sample thickness and temperature. The present study strongly suggests that the thickness dependence reported here is not a consequence of a varying coupling strength or the signature of a different CDW phase, but indeed due to the Fermi surface topology, dimensional crossover and quantum confinement.

**Methods**

1$T$-VSe$_2$ single crystals were grown by chemical vapor transport using I$_2$ as a transport agent and then mechanically exfoliated in-situ (3·10$^{-8}$ mbar) onto Au(111) single crystal substrates. Prior to the exfoliation, the Au(111) surface was cleaned and reconstructed by repeated cycles of Ar$^+$ ion sputtering and annealing at 450°C in ultra-high vacuum (UHV). The exfoliated flakes are hardly visible by optical means in our UHV scanning tunneling microscope (STM) setup. We thus relied on a suitable coverage density produced by our bespoke in-situ exfoliation mechanism to position the STM tip over an exfoliated flake in a systematic scan and search procedure. The STM experiments were done in UHV (base pressure 2·10$^{-10}$ mbar) using tips electrochemically etched from an annealed tungsten wire. The bias voltage was applied to the sample. Tunneling $I(V)$ and differential conductance $dI/dV(V)$ spectra were acquired simultaneously using a standard lock-in technique with a 7.1 mV rms bias modulation at 337.7 Hz.

**Acknowledgement**

We acknowledge A. Morpurgo, Ch. Berthod and T. Giamarchi for stimulating discussions, L. Musy for his contributions to the initial attempts of in-situ exfoliation, and G. Manfrini and A. Guipet for their technical assistance. This project was supported by the Swiss National Science Foundation through Sinergia project (grant 147607).

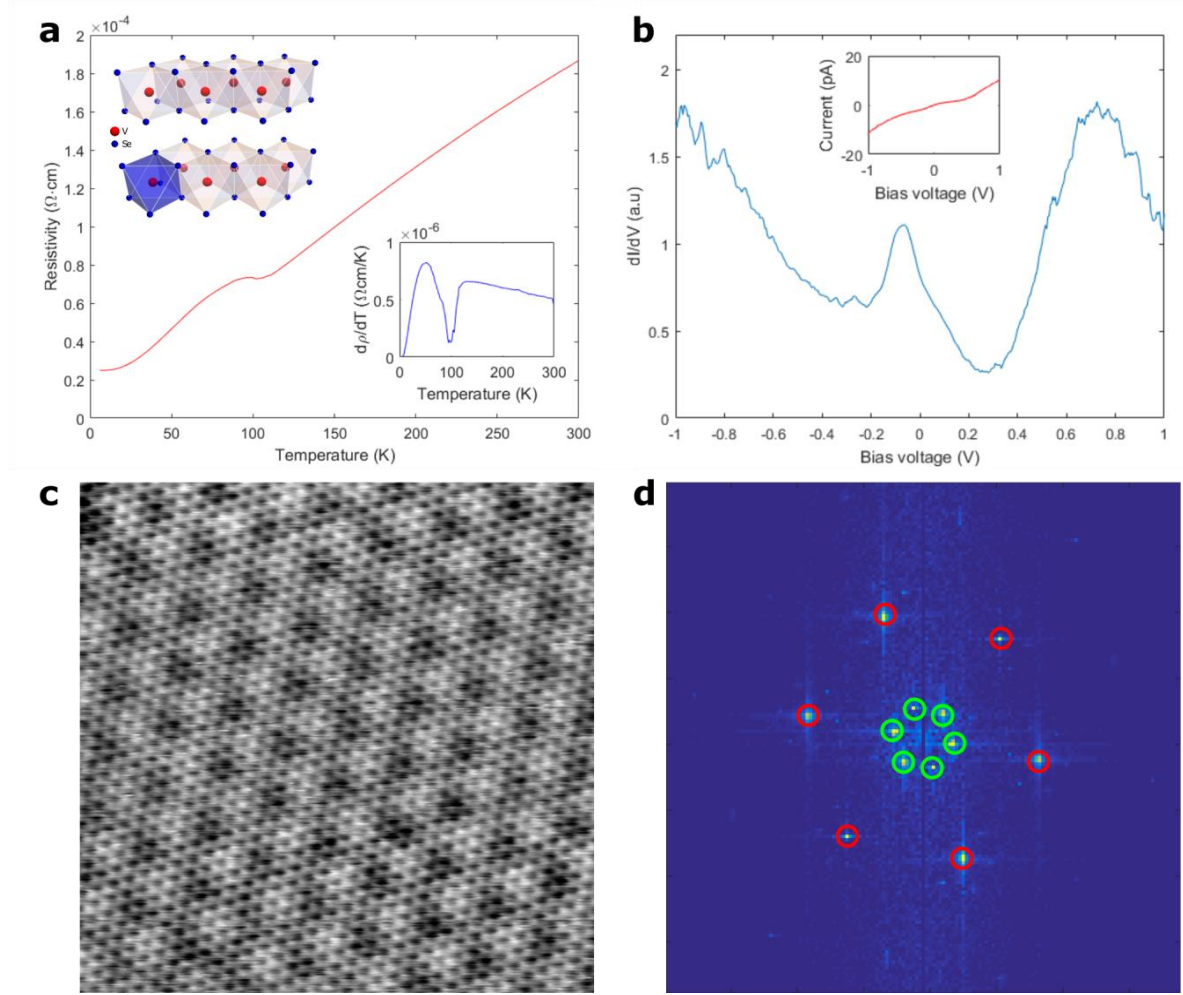

**Figure 1 | Transport and STM characterization of bulk 1*T*-VSe$_2$ single crystals**. (a) Resistivity as a function of temperature with a kink near 105 K associated with the CDW phase transition (left inset: 1*T*-VSe$_2$ crystal structure; right inset: $dR(T)/dT$). (b) Tunnelling spectra measured at 77.6 K on the surface shown in panel c. (c) 10×10 nm$^2$ atomic resolution STM image ($V_{bias}$=-100 meV, $I_t$=10 pA) of a cleaved surface at 77.6 K and (d) corresponding Fourier transform. Red and green circles indicate the first-order atomic lattice and CDW modulation peaks, respectively. The circle size depicts the integration area around each peak used to calculate the CDW order parameter $\psi$.

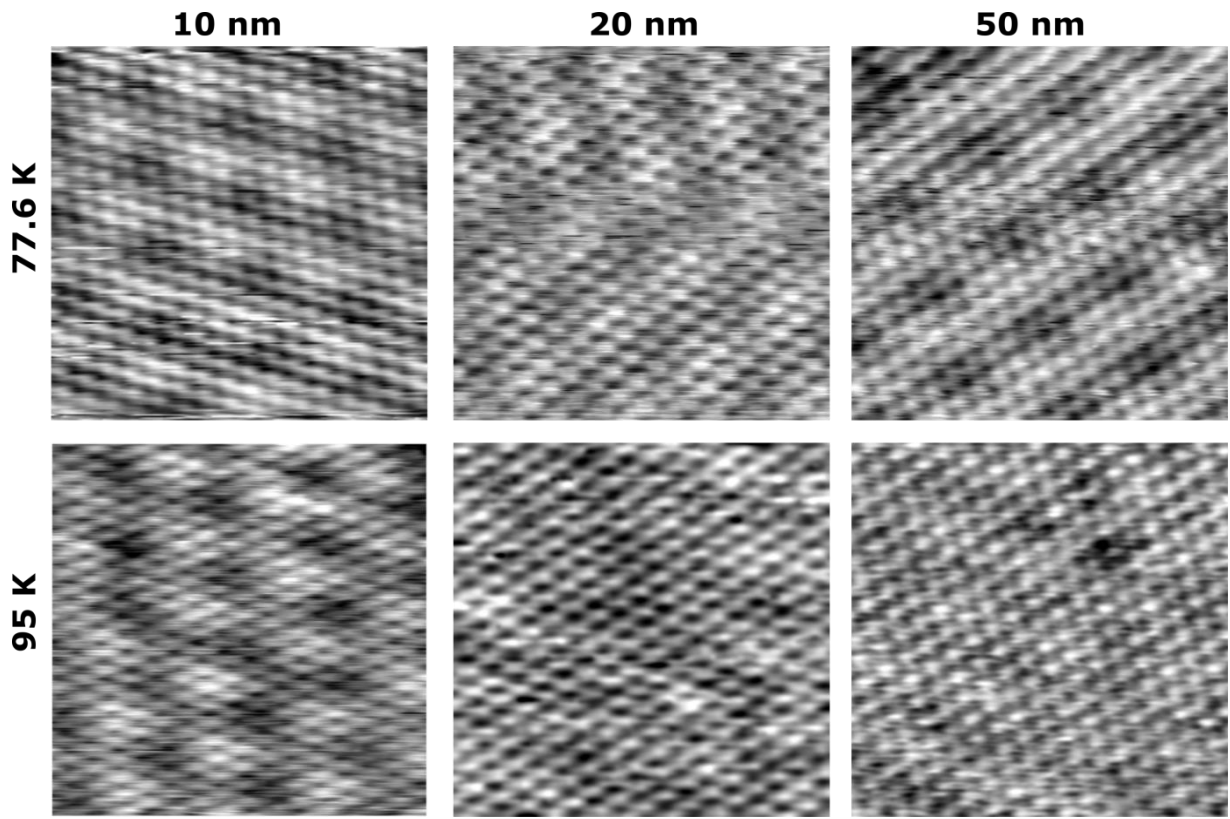

**Figure 2 | STM images of exfoliated 1$T$-VSe$_2$ flakes**. 5×5 nm$^2$ atomic resolution micrographs ($V_{bias}$=-100 meV, $I_t$ =10 pA) measured on different thickness terraces and flakes at 77.6 K and 95.0 K. The atomic lattice is well resolved in all images. The CDW is strongest in the thinnest regions (10 nm) at both temperatures and nearly absent in the 20 nm and 50 nm thin regions at 95.0 K.

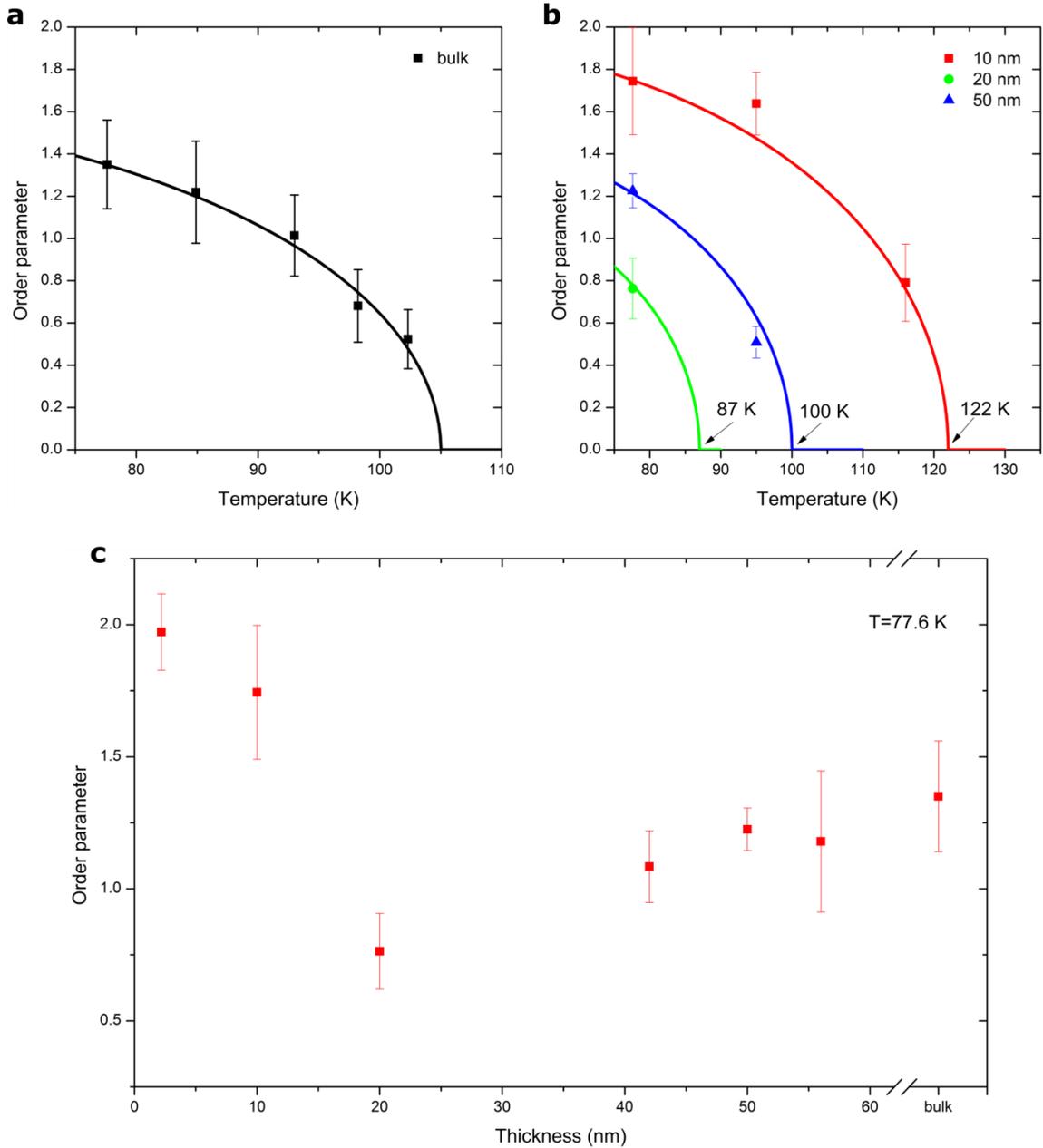

**Figure 3 | Temperature and thickness dependencies of the CDW order parameter $\psi$ in 1T-VSe$_2$.** (a) $\psi$ as a function of temperature in a bulk single crystal near the phase transition. The solid line is a fit to the BCS approximate form $\psi(T) = A \cdot T_c \cdot \tanh\left(1.74 \cdot \sqrt{(T_c/T - 1)}\right)$, where $T_c = 105$ K and the only fitting parameter is determined to be $A = 0.0165 \pm 0.0004$. (b) $\psi$ as a function of temperature for three different thicknesses. The solid lines are calculated with the above BCS interpolation using the bulk scaling factor $A$ and the local $T_c$ calculated from $\psi(T = 77.6\,K)$ for each thickness. (c) $\psi$ as a function of thickness at 77.6 K. In all panels the error bars correspond to the dispersion of $\psi$ as determined by analyzing many different STM images for a given temperature and thickness.

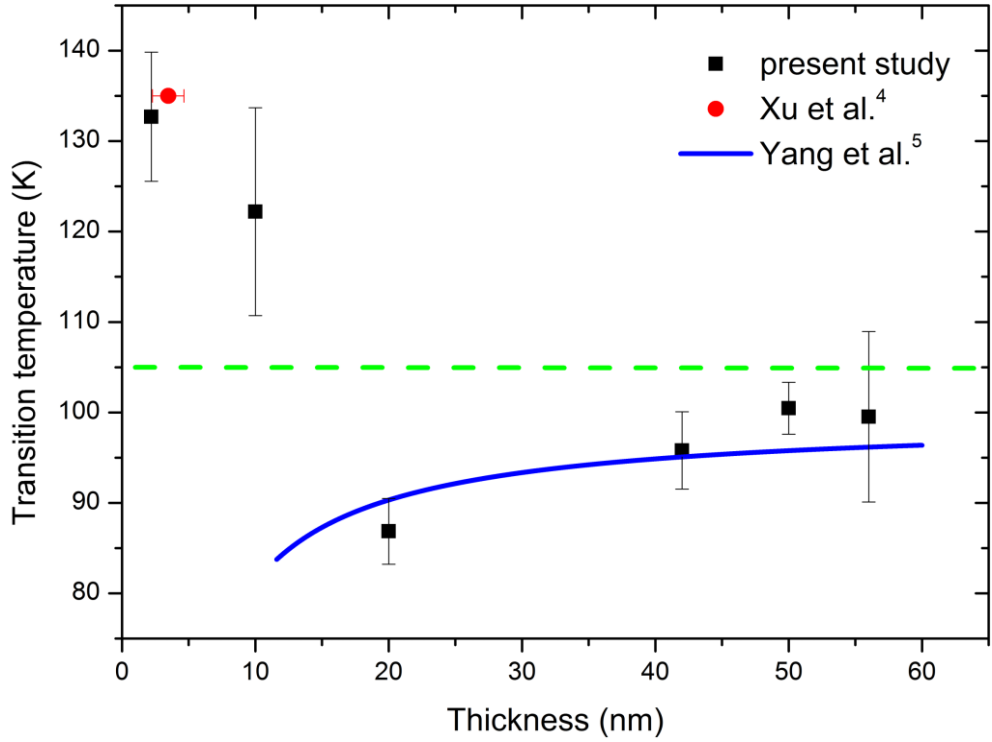

**Figure 4 | Thickness dependence of the CDW transition temperature $T_c$ in 1$T$-VSe$_2$.** Solid squares represent $T_c$ calculated from the charge modulation amplitude imaged by STM using equations (1) and (2). They reveal a clear non-monotonic dependence of $T_c$ on thickness. The vertical error bars correspond to the dispersion from many different STM images. The green dashed line symbolizes the transition temperature of a bulk sample. Our data are in quantitative agreement with and reconcile data from previous transport experiments covering separate thickness ranges (represented by the solid red circle[4] and the solid blue line[5]).